# OptImatch: Semantic Web System with Knowledge Base for Query Performance Problem Determination

Guilherme Damasio
UOIT, Faculty of Science, Computer Science
and IBM Centre for Advanced Studies Toronto
guilherme.fetterdamasio@uoit.ca

Jaroslaw Szlichta
UOIT, Faculty of Science, Computer Science
and IBM Centre for Advanced Studies Toronto
jaroslaw.szlichta@uoit.ca

Piotr Mierzejewski
IBM Canada Ltd
piotrm@ca.ibm.com

Calisto Zuzarte
IBM Canada Ltd
calisto@ca.ibm.com

## ABSTRACT
Database query performance problem determination is often performed by analyzing query execution plans (QEPs) in addition to other performance data. As the query workloads that organizations run, have become larger and more complex, analyzing QEPs manually even by experts has become a very time consuming and cumbersome task. Most performance diagnostic tools help with identifying problematic queries and most query tuning tools address a limited number of known problems and recommendations. We present the OptImatch system that offers a way to (a) look for varied user defined problem patterns in QEPs and (b) automatically get recommendations from an expert provided and user customizable knowledge base. Existing approaches do not provide the ability to perform workload analysis with flexible user defined patterns, as they lack the ability to impose a proper structure on QEPs. We introduce a novel semantic web system that allows a relatively naive user to search for arbitrary patterns and to get solution recommendations stored in a knowledge base either by experts or added by the user tailored to the environment in which they operate. Our methodology includes transforming a QEP into an RDF graph and transforming a GUI based user-defined pattern into a SPARQL query through handlers. The SPARQL query is matched against the abstracted RDF graph, and any matched portion of the abstracted RDF graph is relayed back to the user. With the knowledge base, the OptImatch system automatically scans and matches interesting stored patterns in a statistical way as appropriate and returns the corresponding recommendations. Although the knowledge base patterns and solution recommendations are not in the context of the user supplied QEPs, the context is adapted automatically through the handler tagging interface. We test the performance and scalability of our framework to demonstrate its efficiency using a real query workload. We also perform a user study to quantify the benefits of the approach in terms of precision and time compared to manually searching for patterns.

## Categories and Subject Descriptors
H.2.4 [**Database Management**]: Systems – Query Processing

## 1. INTRODUCTION
### 1.1 Background and Motivation
Much of the world's high-valuable data remain in relational databases (e.g., operational databases and data warehouses [10]). Access to this data is gained through relational query languages such as Structured Query Language (SQL). Complex analytic queries on large data warehouse system are not only done as weekend or end of period canned batch reports. Ad hoc complex queries are increasingly run as part of business operations. As such it is critical to pay attention to performance of these queries.

Database systems themselves are certainly increasingly becoming more sophisticated and able to automatically tune the environments they operate in. General query performance problem determination tools [21], [22] also offer an automated way to database administrators to analyze performance issues that neither requires mastery of an optimizer, nor deep knowledge about the query execution plans (QEPs). However, due to the complexity while the general approach has merit, there is a lack of customization and many refinements are needed, so that the problem determination and tuning process can be truly effective and consumable by the general end-user. Given the specific circumstances and limitations of existing tools, performance analysis today is often best done by manually analyzing optimizer QEPs that provide detail of how queries are executed. Manually analyzing these QEPs can be very demanding and often requires deep expertise particularly with complex queries that are often seen in data warehouse environments. Very often the end users and database administrators resign themselves to opening problem reports to the database vendors so that experts who are well versed in both SQL and analyzing optimizer QEPs can provide recommendations. This can be a time consuming exercise and does not scale well.

```
              19860.9
              NLJOIN
              (    2)
              16246.59
              4909.624
        /---------+---------\
       19.12                 4043
       FETCH                TBSCAN
       (    3)              (    5)
       26.0884               15771
        2.624                 4907
      /---+---\                 |
    19.12    1228             812130
   IXSCAN  SALES_FACT         CUST_DIM
   (    4)    Q2                Q1
   11708.7
    5250
     |
  9.18948e+07
     IDX1
      Q2
```

**Figure 1 Query with NLJOIN**

Existing tools such as IBM® Optim Query Tuner® and IBM Optim Workload Tuner® provide tuning recommendations for specific known problems. While very effective, they do not, however, provide the ability to perform query performance problem determination with flexible user-defined patterns (examples listed below). This is mainly because these tools are agnostic to the complex structure of QEPs. There does not exist a general purpose automated system that would allow for interactive analysis and diagnosis of performance problems by searching for arbitrary patterns within a large number of QEPs. A user not so experienced with QEPs may want to answer simple questions. For example, after searching and determining the cost of a table scan on a particular table, the user may want to know how many queries in the workload do an index scan access on the table and get a sense of the implications of dropping the index by comparing the index access cost to that of the table scan. Even with more experienced database administrators, often there are clues from monitoring data that provide hints of certain characteristics of QEPs that are not easily found by using typical search tools like *grep*. For instance, given a large number of queries, say 1000 queries, and the corresponding workload QEPs:

- Find all the queries in the workload that might have a spilling hash join below an aggregation and the cost is more than a constant *N*.
- Find all the subqueries that have a cost that is more than 50% of the total cost of the query and provide details of the subquery operators (name, cost, and input operators).
- Find all the queries that have an outer join involving the same table somewhere in the plan below both sides of a hash join.
- Perform cost based clustering and correlate results of applying expert patterns to each cluster.

We consider making it easier and faster to automatically answer questions like the above in our work. We provide a flexible system OptImatch that performs analysis over large and complex query workloads, in order to help diagnosis optimizer problems and retrieve solutions that were previously provided by experts. The optImatch system drastically lowers the skill level required for optimizer access plan problem determination through advanced automated pattern matching and retrieving of solution recommendations of previously discovered performance problems for single queries and large query workloads. OptImatch is very well received and is proving to be very valuable in the IBM support of business clients and database optimizer development organization.

At the enterprise level, major commercial relational database systems such as IBM DB2®, Oracle®, and Microsoft® SQL Server® are deployed in environments where finding all available optimizations and performance tuning strategies becomes necessary to maintain the usability of the database. Traditional optimization methods often fail to apply when logical subtleties in queries and database schemas circumvent them. The examples of this include cases, where the recommended performance enhancement is to index a table in a particular way, prescribe an integrity constraint such as functional dependency [14] or order dependency [15], create a materialized view [5] or to rewrite manually the proposed SQL query, where orthogonal approach with machine optimization [4], [19], [23] failed to rewrite the query to get the same answer but with a better performance.

The problem pattern comprises a list of operators having particular properties that are of interest to a user, as exemplified in some of the aforementioned problems. By incorporating our query performance problem determination system many optimization problems could be automatically identified and resolved. Figure 1 depicts an example of a text graph version of a snippet of a QEP from IBM DB2. The snippet shows a nested loop join (NLJOIN) of the SALES_FACT table accessed using an index scan (IXSCAN) with other columns fetched (FETCH) from the table and then joined to the CUST_DIM table. The numbers immediately above the operator or table name show the estimated number of rows flowing out (cardinality). The numbers in parenthesis show the operator number. Operators are also referred to as Plan OPerators (pop) or LOw LEvel Plan OPerators (LOLEPOP) in this paper. Each operator has an estimated Input/Output (I/O) cost, the bottom number below the operator number, and a cumulative cost for itself and all operator below it, the number immediately below the operator number. In the depicted example, a user could be concerned with NLJOIN that has an inner stream of type table scan (TBSCAN). Such query is costly as the NLJOIN operator scans the entire inner table CUST_DIM for each of the rows from the outer SALES_FACT table. An example of a solution recommendation might be to provide a recommendation to create an index of the target table of the TBSCAN, in this case CUST_DIM.

In recent years, more and more customer queries are generated automatically by query managers (such as IBM Cognos®) with business users providing only specific parameters through graphical interfaces [7], [8]. Specific parameters are then automatically translated by query managers into executable SQL queries. Based on analyzing IBM customer workloads there is essentially no limit to the length of the query generated automatically by query managers. It is quite usual to find queries with over one thousand lines of SQL code (hundreds of operators). Such queries are very complex and time consuming to analyze with nesting and stitching of several subqueries into a larger query being a common characteristic. Another common feature is repetitiveness, where similar (or even identical) expressions appear in several different parts of the same query, for instance, in the queries referring to the same view or nested query block multiple times [13], [20]. If there is need to improve the performance of such complex queries, when optimizer failed, it could be time consuming to do this manually. It could take hours or even days to analyze a large query workload. Our goal is automate this process as much as possible, and therefore save significant amount of time spent by users on query performance problem determination.

The OptImatch system makes this process easier. We decided to use the RDF format as it allows one to easily retrieve information with the SPARQL query language. SPARQL has the capability for querying optional and required graph patterns. Another powerful feature exploited in SPARQL is that of property paths. A property path is a possible route through a graph between two graph nodes. SPARQL property paths provide a succinct way to write parts of graph patterns and to also extend matching patterns to arbitrary length paths. With property paths, we can handle recursive queries, for instance, search for a descendant operator that does not necessarily have an immediate relationship (connection) with its parent. We can also search for patterns that appear multiple times in the same QEP. Last but not least, SPARQL allows graph traversal and pattern matching in a very efficient way [3], enabling analysis of a large number of complex QEPs in a short period of time.

While the focus of this work is on query performance problem determination, our methodology can be applied to other general software problem determination [24], assuming that there exists automatically or dynamically generated diagnostic information than needs to be further analyzed by an expert. Broadly, the contemplated diagnostic data may be human-readable and intended for review by human users of the system to which the diagnostic data relates. Examples of possible diagnostic data include log data relating to network usage, security, or compiling software, as well as software debug data or sensor data relating to some physical external system. In these scenarios, the problem pattern may correspond to any sequence of data points or interrelationships of data points that are of diagnostic interest.

## 1.2 Contributions

The main contributions of this paper appear in Section 2 and Section 3 as follows.

1. We developed a semantic web tool to transform a QEP into an abstracted artefact structure (RDF graph). We propose in our framework to model features of the QEP into a set of entities containing properties with relationships established between them. (Section 2.1)
2. We provide a web-based graphical interface for the user to describe a problem pattern (pattern builder). The tool transforms this pattern into a SPARQL query through handlers. Handlers provide the functionality of automatically generating variable names used as part of the SPARQL query. The SPARQL query is executed against the abstracted RDF structure and any matched portions of RDF structure are relayed back to the user. We present a suite of real-world IBM customer problem patterns that illustrate the issues related to query performance, which are then used in Section 3 for experimental performance evaluation. (Section 2.2).
3. We added a knowledge base capability within the tool that could be populated with some expert provided patterns and solution recommendations as well as allow users to add their own patterns and recommendations. The system automatically matches problem patterns in knowledge base to the QEPs and if there are any search results ranks them using statistical correlation analysis. OptImatch distinguishes between a pattern builder and a tagging handler interface to achieve generality and extensibility. In a nutshell, the pattern builder allows the users to specify what is wrong with the query execution plan (static semantics), and the handler tagging interface defines how to report and fix it (dynamic semantics) through automatically adopting the context. Since the knowledge base patterns and solution recommendations are not in the context of the user supplied QEPs, we have defined the language that users can use to add dynamic context to the recommendations. (Section 2.3)
4. An experimental evaluation showing the performance and effectiveness of our techniques was carried out using real IBM customer datasets. We experimented with different problem patterns, and show that our framework runs efficiently over large and complex query workloads. Our performance evaluation reveals that the time needed to compute a search over a specified problem pattern against a QEP increases linearly with the size of the workload, the number of operators in the QEP and the number of pattern/recommendations in the knowledge base. Finally, we show through a user study that our system is able to save a significant amount of time to analyze QEPs. Moreover, we quantify in the user study the benefits of our approach in terms of precision over manual pattern searching. (Section 3)

In Section 4, we discuss related work. We conclude and consider future work in Section 5.

To the best of our knowledge, we are the first to provide a system for query performance problem determination by applying QEP feature transformation through RDF and SPARQL. This work we feel opens exciting venues for future work to develop a powerful new family of problem determination techniques over existing optimizer performance analysis tools and other diagnostic data exploiting graph databases.

## 2. SYSTEM
## 2.1 Transforming Diagnostic Data

Even though optimizer diagnostic data may differ in some ways between various database management systems, their major characteristics remain the same. Query performance diagnostic information is usually in the form of QEPs formatted in readable text form. An example of the portion of the QEP generated by the IBM DB2 database engine is presented in Figure 1. A QEP includes diagnostic information about base objects (e.g., tables, views and indexes), operators (e.g., join, sort and group-by) as well as costs and characteristics associated with each operator.

Some properties of operators are included in a QEP in the tree diagram as in Figure 1 (e.g., cardinality total cost, Input/Output cost, cumulative cost), wherein other properties appear as separate textual blocks identified by operator number (e.g., cumulative CPU cost, cumulative first row cost and estimated bufferpool buffers). Furthermore, some properties are common between different types of operators (e.g., cardinality, total cost and CPU cost), while others are specific to certain operators. For instance, NLJOIN has a property fetch max, and TBSCAN has a property max pages, but not vice versa. A QEP also contains some other detailed diagnostic data, including information about the DBMS instance and environment settings. All of the techniques described in this paper have been implemented given IBM DB2 QEPs. Hence, much of the discussion through the rest of the paper is framed in the terminology and characteristics of IBM DB2. However, the techniques that are described have general applicability, and can be used with any other DBMS product or other diagnostic data that lends itself to property graph representation.

A QEP can be viewed as a directed graph that indicates the flow of operations processing data within the plan. QEPs resemble a tree

```
<http://explainPlan/PlanPop/1> <http://explainPlan/PlanPred/hasPopType> "NLJOIN" .
<http://explainPlan/PlanPop/1> <http://explainPlan/PlanPred/hasOuterInputStream> <http://explainPlan/PlanPop/3> .
<http://explainPlan/PlanPop/1> <http://explainPlan/PlanPred/hasInnerInputStream> <http://explainPlan/PlanPop/5> .
<http://explainPlan/PlanPop/3> <http://explainPlan/PlanPred/hasPopType> "FETCH" .
<http://explainPlan/PlanPop/3> <http://explainPlan/PlanPred/hasEstimatedCardinality> "19.12" .
<http://explainPlan/PlanPop/5> <http://explainPlan/PlanPred/hasPopType> "TBSCAN" .
<http://explainPlan/PlanPop/5> <http://explainPlan/PlanPred/hasEstimatedCardinality> "4043.0".
<http://explainPlan/PlanPop/5> <http://explainPlan/PlanPred/hasTotalCost> "15771.0" .
<http://explainPlan/PlanPop/5> <http://explainPlan/PlanPred/hasIOCost> "49007.0" .
<http://explainPlan/PlanPop/5> <http://explainPlan/PlanPred/hasJoinInputLeg> "INNER" .
<http://explainPlan/PlanPop/5> <http://explainPlan/PlanPred/hasOutputStream> <http://explainPlan/PlanPop/1> .
<http://explainPlan/PlanPop/5> <http://explainPlan/PlanPred/hasInputStreamPop> <http://explainPlan/PlanBaseObject/CUST_DIM> .
<http://explainPlan/PlanBaseObject/CUST_DIM> <http://explainPlan/PlanPred/hasEstimateCardinality> "812130.0"
```

**Figure 2 Generated RDF in textual representation**

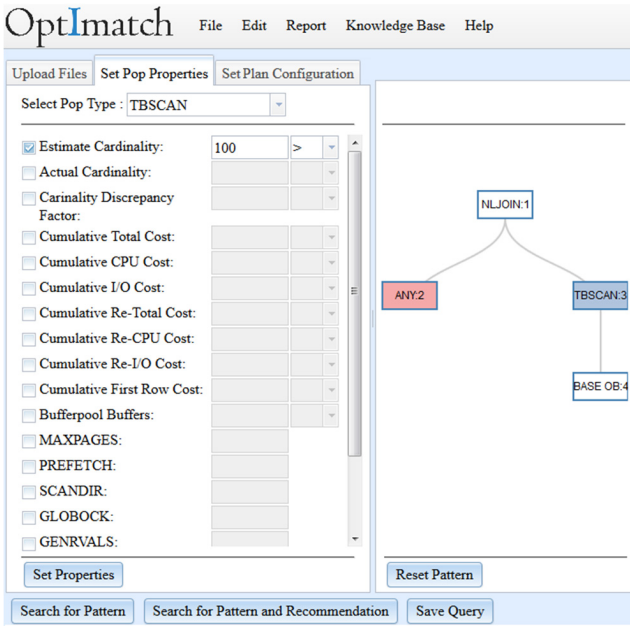

**Figure 3 Web-based Graphical Interface (Pattern Builder)**

structure, where each node (operator) possesses numerous properties and is considered as one of the inputs to a derived ancestor node. LOLEPOPs in a QEP are connected to their parents as inputs streams. These inputs can be identified with three different types 1) outer input (left input of the parent operator) 2) inner input (right input of the parent operator) and 3) general input stream (generic input used for any kind of operator).

The LOLEPOPS may be understood, at the level of abstraction of the DBMS user, as indivisible operations that are directly executed by the DBMS, with each LOLEPOP carrying a stated cost. The stated cost for each LOLEPOP represents an estimate of server resources, generated by the DBMS system based on a proposed SQL query by taking into account the particular properties of the database. The overall QEP is machine-generated by the DBMS Optimizer [12]. It is machine-optimized to gravitate towards the lowest total cost LOLEPOPs attainable by the DBMS's optimizer. The plan structure is highly dynamic and can change based on configuration, statistics of the data associated with referenced base objects and other factors even if query characteristics remain similar. However, plan changes are difficult to spot manually as they tend to spawn thousands of lines of informative details for more complex queries in the workload.

RDF is a labeled directed graph built out of triples, each containing subject (resource), predicate (property or relationship) and object (resource or value). RDF does not enforce specific schema, hence, two resources in addition to sharing properties and relationships, can also be described by their own unique predicates. This property of RDF is beneficial to describe and preserve various types of complex diagnostic information about QEPs. Even though RDF inherently does not possess a particular structure, such structure can be enforced by specifying predicates (for example, defining predicates, such as *hasInputStreamPop* or *hasOutputStreamPop*, and *hasInnerInputStreamPop* or hasOuterInputStreamPop and using them to establish relationships between resources (LOLEPOPs)). This allows one to recreate the tree structure and characteristics used in QEPs.

**Algorithm 1** TransformingQEPs
**Input:** query execution plan files *QEPFs*[ ]
**Output:** execution plans represented as RDF Graphs, *RDFGs*[]
1: **forall** *qepf* in *QEPFs*[ ]
2:   *i* := 0
3:   *rdfg* := convert *qepf* into RDF graph model
       with Jena RDF API
4:   *RDFGs*[*i*] := *rdfg*
5:   *i* := *i* + 1
5: **end forall**
6: **return** *RDFGs*[ ]

We propose in our framework to model features of the QEP into a set of entities containing properties with relationships established between them. In these terms, a QEP can be modelled into LOLEPOPs (entities), type, cardinality and costs (properties) and input/output streams (relationships). This model, represented in our framework by means of Apache Jena RDF API, is applied to QEPs provided by the user and persisted in a transformation engine (Algorithm 1). Jena is a Java API which can be used to create and manipulate RDF graphs. Jena has object classes to represent graphs, resources, properties and literals. The result is a transformation of the QEP into an RDF graph, where each LOLEPOP represents an RDF Resource, each property and relationship represents an RDF Predicate and each property value is represented by an RDF Object. During the transformation from the QEP file to the RDF graph additional derived properties can be defined by analyzing resource properties. For instance, the *hasTotalCostIncrease* predicate allows us to calculate and store the total cost of the LOLEPOP by subtracting the cost of the input

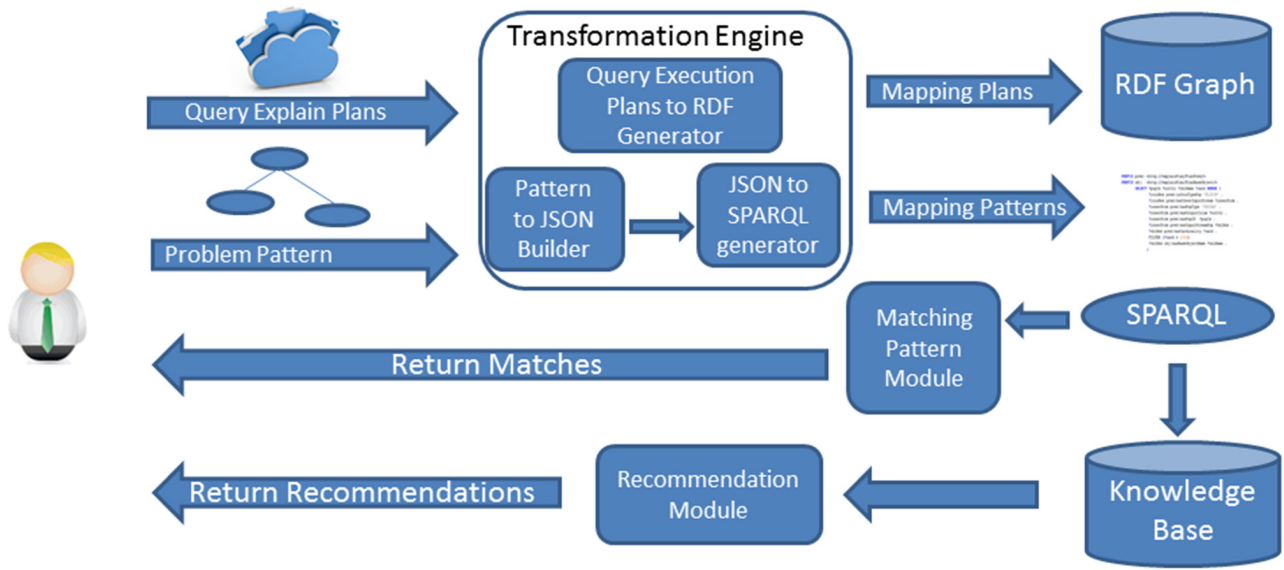

**Figure 4 System Architecture**

LOLEPOPs from currently LOLEPOP being analyzed. The Generated RDF graph can then be preserved in memory ready for analysis by the transformation engine. The OptImatch System architecture is illustrated in Figure 4.

An example of an auto-generated RDF graph in textual representation is presented in Figure 2. Figure 2 depicts an RDF representation of the LOLEPOPs shown in Figure 1. Each RDF statement is in the form of a triplet including a resource, a predicate and an object. In the code presented, statements referring to the resource "http://explainPlan/PlanPop/5" represents LOLEPOP #5. Various predicates are shown, each encoding a piece of information from the QEP. For example, there are predicates that specify LOLEPOP #5's total cost (15771) and estimated cardinality (4043). The RDF representation of all other LOLEPOPs is generated accordingly.

## 2.2 Searching for Problem Patterns

SPARQL is the RDF Query Language. The SPARQL standard is maintained by the W3C. Our system can accomplish the step of searching for user-defined problem patterns in QEPs by transforming patterns into SPARQL queries directed to the abstracted RDF derived from the QEPs. SPARQL performs graph traversal and pattern matching efficiently. This allows one to analyze complex patterns over large query workloads in a short period of time. SPARQL contains capability for querying optional and required patterns of the graph with arbitrary length paths. Furthermore, SPARQL property paths provide a succinct way to match patterns in the RDF graph. This includes recursive queries, such as looking for descendants operators that do not necessarily have an immediate connection with their parent (see Pattern B in Section 2.3), and searching for patterns that appear multiple times in the same query execution plan. While one could consider using any property graph representation framework, RDF was used also for convenience since DB2 supports RDF file format and SPARQL querying across all editions from DB2 10.1, when the RDF specific layer, DB2 RDF Store, was added. The DB2 RDF Store is optimized for graph pattern matching.

Query performance problems can usually be described as problem patterns in the QEP. A problem pattern is a set of optimizer plan features and characteristics specified in a particular order and containing properties with predefined values. Figure 3 displays a web-based graphical user interface (pattern builder) used in our system wherein a user can express the problem pattern by selecting various properties of LOLEPOPs and plan properties that a user might be interested in within the QEP. In the depicted example of problem pattern (**Pattern A**), the user is concerned with a LOLEPOP that: (i) is of type NLJOIN; (ii) has an outer input stream of type any (ANY) with cardinality greater than one (meaning that the outer is likely to be more than one row and consequently the inner will be accessed multiple times; (iii) has an inner input stream of type TBSCAN; and (iv) the inner input stream has large cardinality (greater than 100). The depicted graphical user interface generates an example structure of a LOLEPOP that matches the selected properties. In this case, the described LOLEPOP is a nested loop join operator (NLJOIN) with some operator (ANY) on the outer input stream and a table scan (TBSCAN) on the inner input stream. Such a pattern is costly as deduced by satisfying the cardinality conditions. The NLJOIN operator scans the entire table (TBSCAN) for each of the rows from the outer operator ANY. It would likely be of value for a subject matter expert to spend time and attention to try to optimize queries matching this problem pattern in the QEP. (System recommendations are described in details in Section 2.3.)

**Algorithm 2 TransformingProblemPattern**
**Input:** problem pattern *probPat*
**Output:** problem pattern *probPat* transformed to SPARQL query
1: *probPatJSON* := translate problem pattern *probPat* into JSON Object
2: *sparql* := transform JSON object *probPatJSON* through handlers into SPARQL query
3: **return** *sparql*

When specifying a problem pattern using the graphical user interface (GUI) for generality and flexibility sake, the user can choose between two types of relationships: immediate and descendant. Descendants are operators that are successors but not necessarily immediately below the current LOLEPOP. In that case, the path between the parent and the descendant child is called the cloud. (In the general case, the cloud can contain any number of operators.) For instance, in Figure 1, LOLEPOP #4 is an immediate child of LOLEPOP #3 and LOLEPOP #4 is a descendant child of LOLEPOP #2.

Once the desired problem pattern is defined by the user by describing LOLEPOPs, their characteristics and relationships, it is then automatically translated (Algorithm 2) into a JavaScript Object Notation (JSON). This object is constructed to contain a transformation of the properties specified in the pattern builder to the RDF resources and the predicates defined in the model used in the QEP. In Figure 5, we present an example JSON Object that contains properties specified in the pattern builder (Figure 3). The generated JSON Object is an array of objects describing each resource operator and its relationships. For instance, the portion of JSON Object describing LOLEPOP with ID 1 has specified type NLJOIN, an estimated cardinality value of more than 100 and relationship with two immediate children operators, LOLEPOP with ID 2 and LOLEPOP with ID 3.

```
{"pops":[
{"ID":1,"type":"NLJOIN","popProperties":
    [ {"id":"hasEstimateCardinality","value":"100","sign":">"},
      {"id":"hasOuterInputStream","value":2,"sign":"Immediate
            Child"},
      {"id":"hasInnerInputStream","value":3,"sign":"Immediate
            Child"}]},
{"ID":2,"type":"ANY","popProperties":
      [{"id":"hasOutputStream","value":1}]},
{"ID":3,"type":"TBSCAN","popProperties":
      [{"id":"hasInputStream","value":4,"sign":"Immediate
      Child"},
        {"id":"hasOutputStream","value":1}]},
{"ID":4,"type":"BASE
 OB","popProperties":[{"id":"hasOutputStream","value":3}]}]},
        {"planDetails":[]}
```

**Figure 5 JSON Object**

The transformation engine uses JSON Objects to auto-generate an executable SPARQL query. An example of the autogenerated SPARQL query is presented in Figure 6. The URIs broadly match the RDF graph generated based on the QEP in Figure 1, and various SPARQL query operators and operands match the elements of the problem pattern indicated by the user.

An autogenerated SPARQL query is composed of two main parts, the SELECT clause that defines variables that appear in the query results, and the WHERE clause that defines resource properties that should be matched against the specified RDF graph. The variables that appear in query results are specified by prefixing variable name with "?" symbol, i.e., "?variable_name" (and can be referenced multiple times in the WHERE clause). The same convention is used to define variables to establish relationship between resources and the ones used to filter retrieved resources.

Our framework allows us to autogenerate SPARQL queries with a wide range of characteristics, including nesting, filtering, multiple resource mapping, and specifying property paths as well as blank nodes. Blank nodes in RDF indicate the existence of unnamed or previously undefined resources. We introduce the concept of *handlers* to facilitate this. Handlers provide the functionality of automatically generated variable names used for the retrieval of query results, filtering of retrieved values, and establishing relationships between resources and blank nodes.

Handler generation is performed in a modular manner, by building the SPARQL query one layer (one operator) at the time over portions of JSON Object. In order to generate the SPARQL query, we define four types of handler variables: *result handlers, internal handlers, relationship handlers* and *blank node handlers*. Result handlers are created based on identifiers (sequential identifiers assigned to each LOLEPOP as shown in the graphical user interface in Figure 3), i.e., ?*pop1* and ?*pop2* etc. For instance, in our SPARQL query, the result handler ?pop1 is a resource returned to the user, and is also used in the WHERE clause to identify this resource as NLJOIN by adding the predicate *hasPopType*.

```
PREFIX popURI: <http://explainPlan/PlanPop/>
PREFIX baseObjURI: <http://explainPlan/PlanBaseObject/>
PREFIX predURI: <http://explainPlan/PlanPred/>
PREFIX planURI: <http://explainPlan/PlanDetails/>
  SELECT (?pop1 AS ?TOP) (?pop2 AS ?ANY2)
          (?pop4 AS ?BASE4)
  WHERE {
    ?pop1 predURI:hasPopType "NLJOIN" .
    ?pop1 predURI:hasEstimateCardinality ?internalHandler1 .
        FILTER ( ?internalHandler1 > 100) .
    ?pop1 predURI:hasOuterInputStream
        ?BNodeOfpop2_to_pop1 .
    ?BNodeOfpop2_to_pop1 predURI:hasOuterInputStream
        ?pop2 .
    ?pop2 predURI:hasOutputStream ?BNodeOfpop2_to_pop1.
    ?BNodeOfpop2_to_pop1 predURI:hasOutputStream ?pop1 .
    ?pop1 predURI:hasInnerInputStream
        ?BNodeOfpop3_to_pop1.
    ?BNodeOfpop3_to_pop1 predURI:hasInnerInputStream
        ?pop3 .
    ?pop3 predURI:hasOutputStream ?BNodeOfpop3_to_pop1.
    ?BNodeOfpop3_to_pop1 predURI:hasOutputStream ?pop1 .
    ?pop3 predURI:hasPopType "TBSCAN" .
    ?pop3 predURI:hasInputStream ?BNodeOfpop4_to_pop3 .
    ?BNodeOfpop4_to_pop3 predURI:hasInputStream ?pop4 .
    ?pop4 predURI:hasOutputStream ?BNodeOfpop4_to_pop3.
    ?BNodeOfpop4_to_pop3 predURI:hasOutputStream?pop3 .
    ?pop4 predURI:isABaseObj ?internalHandler2 .
} ORDER BY ?pop1
```

**Figure 6 Autogenerated SPARQL Executable Query**

Internal handlers are used to filter results. Identifiers of internal handlers are not tied to a specific resource. Their identifiers are automatically incremented on the server. For instance, the handler ?*internalHandler1* is generated to provide the filtering of cardinality property by first associating it with ?*pop1* (?pop1 predURI:hasEstimatedCardinality ?internalHandler1) and then utilizing it in the FILTER clause (FILTER (?internalHandler1 > 100)).

Relationship handlers establish connection between resources based on information about the hierarchy of operators retrieved from the JSON Object (e.g., {"id": "hasOuterInputStream","value":

2,"sign": "Immediate Child"}). The relationship handlers are used in conjunction with blank node handlers to resolve ambiguity problems. Ambiguity problems are encountered when the same LOLEPOP is absorbed in the different parts of the QEP. Such a LOLEPOP, for example, a common sub expression with a temporary table (TEMP) that has multiple consumers, has the same cardinality in all the consumers which may produce different results. This might be the case, for example, when a common sub expression TEMP is consumed by both a NLJOIN and a HSJOIN in the different parts of the QEP applying different predicates. In such a case, the output columns of NLJOIN and HSJOIN might differ even though the input common sub expression TEMP into each of them is the same. In the above example, ?*pop1* resource has the predicate *hasOuterInputStream* connecting it to ?*pop2* via the blank node ?*BnodeOfPop2_to_pop1* (?pop1 predURI: hasOuterInputStream ?BNodeOfpop2_to_pop1). This design ensures the uniqueness of each resource instance in the QEP.

The autogenerated SPARQL query through handlers is matched against the abstracted RDF structure containing information about the QEP. It maps any matched portions of the abstracted RDF structure back to the corresponding diagnostic data (Algorithm 3). Figure 1 represents an example of the DBMS QEP that contains problem pattern specified in Figure 3.

**Algorithm 3** FindingMatches
**Input:** problem pattern *probPat*,
 query execution plan files *QEPFs*[ ]
**Output:** matches found in query execution plans
1: *RDFGs*[ ] := TransformingQEPs(*QEPFs*[ ])
2: *sparql* := TransformingProblemPattern(*probPat*)
3: **forall** *rdfg* in *RDFGs*[]
4:  *matchProbPat*[ ] := match abstracted problem pattern *sparql*
   against query execution plan *rdfg*
5:  **if** (*matchProbPat* != *empty*)
6:   *matchProbPatDet*[ ] := detransformation by relating
    any matched portions of RDF structure
    *matchProbPat* back to corresponding query plan
7:   *MATCHES*[ ].append(matchProbPatDet[ ])
8:  **endif**
9: **end forall**
10: return *MATCHES*[ ]

Matching problem patterns against diagnostic data allows for dynamic analysis of ad-hoc patterns. However, beyond single pattern matching, the tool usage can vary from problem identification and analysis to solution recommendations as described in the following section.

## 2.3 Finding Solutions in Knowledge Base

The OptImatch system has the ability to access the knowledge base to provide solutions to the known problems. The knowledge base is populated with predetermined problem patterns and associated query plan recommendations by subject matter experts (e.g. IBM employees or expert database administrators). OptImatch promotes and supports collaboration among developers, experts and database administrators to create library of patterns and recommendations.

Once defined, the problem pattern is preserved in the knowledge base in two forms: an executable SPARQL query that is applied to the QEP provided by the user and as an RDF structure describing this pattern. Although the knowledge base problem patterns and solution recommendations are not in the context of the user supplied QEPs, the context for problem patterns is adapted automatically through the handlers tagging with the defined language.

Once a problem pattern to be stored in the knowledge base is described by an expert, it is translated into the SPARQL query that includes result handlers (Section 2.2). The result handlers can have *aliases* associated with them. Looking at the example SPARQL snippet we can see that the result handler ?pop1 has been assigned an alias ?TOP and ?pop4 an alias ?BASE4. These aliases are used to tag the recommendation to the specified result handlers. Tagging allows for identifying a specific result handler or a set of result handlers to be returned. This allows OptImatch to list table names, column names and predicates etc., in the context of the QEP provided by the user even though these are not available when the recommendations were created.

**Algorithm 4** SaveRecommendationsKB
**Input:** problem patter *probPat*
 suggested recommendations *recomms*[]
 *current knowledge base KB*[ ]
**Output:** updated knowledge base KB[ ]
1: *sparql* := TransformingProblemPattern(*probPat*)
2: save abstracted problem *sparql*, problem pattern represented
  as RDF and corresponding recommendations *recomms*[]
  in knowledge base KB[ ] with handlers tagging interface
3: **return** *KB*[ ]

Our language allows for surrounding static parts of recommendations with dynamic components generated through aliases by preceding each alias of the handler with an "@" sign. This approach is also used to limit the number of resource handlers returned to the user since in complex queries there can be large number or result handlers generated, however, only some of them might be significant to the recommendation.

**Algorithm 5** FindingRecommendationsKB
**Input:** query execution plan files *QEPFs*[ ]
 knowledge base *KB*[ ]
**Output:** solution recommendations for queries that match
 *QEPFs*[]
1: **forall** *qepf* in *QEPFs*[]
2:  *queryReccomendation*[ ] := match specified qepf against
  knowledge base *KB*[] using statistical analysis and
  provide recommendations to diagnostic data through tags
  of handlers
3:  **if** (*queryReccomendation* != *empty*)
4:   *queryRecommendations*[ ].
    append(queryReccomendation)
5:  **else**
6:   *queryRecommendations*[ ].
    append("There is currently no recommendation in
     knowledge base")
7:  **endif**
8: **end forall**
9: **return** *queryRecommendations*[]

When a user wants to include multiple result handlers and apply the same rules to each of them, it can be accomplished using array brackets e.g., [@TOP, @ANY2]. For common patterns (appearing multiple times in the same QEP) a user may limit the number of occurrences of the pattern that is returned in recommendation results. In the following example, [@TOP, @ANY2]:1, only the first occurrence of @TOP and @ANY2 is returned and the specifics of the LOLEPOP types and names are obtained from the context of each occurrence.

Furthermore, a user can make use of various *helper functions* constructed to allow for interactions with base tables, indexes and materialized query tables (MQTs). These functions provide means

```
                          0.157686
                           NLJOIN
                          (    5)
                           644901
                           751020
                  /--------+---------\
          8.78417e+06                 1.79511e-08
            >HSJOIN                      TBSCAN
            (    6)                     (   13)
             633711                     2267.08
             750436                     583.334
          /---+----\                       |
   78417e+06   5.99144e+06              0.174681
    ^HSJOIN      TBSCAN                   TEMP
    (    7)     (   12)                 (   14)
     561520      68023.4                 2267.07
     664808      85628                   583.334
                   |                       |
              5.99144e+06                0.174681
            TELEPHONE_DETAIL              >NLJOIN
                 Q1                     (   15)
                                         2267.07
                                         583.334
```

**Figure 7 Query with Left Outer Join**

to list column predicates and table names specific to each occurrence of the pattern in the context of the user provided query execution plan. For instance, a following expression @TOP.listColumns("PREDICATE") lists columns from an alias handler in the predicate indicated by the keyword PREDICATE.

An expert can also use ?TOP alias tagging handler to indicate that when such pattern is encountered all input columns (using keyword INPUT) coming from ?BASE4 object into the NLJOIN should be listed and are valid candidates for the index creation. This can be accomplished by tagging recommendation with following expression:

"*Create index on table @BASE4 on columns @TOP.listColumns("INPUT")*",

and adding it to the knowledge base with the corresponding pattern.

Our system can look through all the QEPs supplied and iterate through both the user-defined problem patterns and the library of expert provided patterns with corresponding recommendations. If there is a match between the problem pattern in the knowledge base and the QEP, one or more query plan recommendations are returned with the appropriate context.

Our system returns ranked recommendations by using statistical correlation analysis. QEPs typically have operators, estimated or actual cost, frequency or priority metrics associated with them (as described in more details in Section 2.1). These characteristics are critical to the database system in terms of performance. Based on these characteristics a prioritized list of recommendations is provided by the system. The ranked recommendations are provided with a confidence score. For instance, in the example described in Section 2.2 with NLJOIN, the query plan problem determination program could output the recommendation (by automatically generating context) to create an index of the CUST_DIM table that is the source for the TBSCAN, as this could be the recommendation stored in the knowledge base created by the experts. An example of the syntax for creating index is illustrated in the previous paragraph.

```
                          1.311e-08
                           IXSCAN
                          (   38)
                           16.9825
                              3
                              |
                         2.55276e+08
                            IDX9
                          TRAN_BASE
                            Q21
```

**Figure 8 Estimation of the execution cost**

An alternate recommendation may be to collect column group statistics in order to get better cardinality estimates so that the optimizer may choose a hash join instead of a nested loop join. Ranking between these two recommendations can be aided with statistical correlation analysis comparing the QEP context of cardinality and cost estimates with that in the expert provided patterns.

OptImatch can provide advanced guidance with a variety of recommendations for example, changing database configuration, improving statistics quality, recommending materialized views, suggesting alternate query and schema design changes, and recommending integrity constraints that promote performance. We illustrate some examples of these below.

As an example of a problem related to query rewrite, we describe the pattern that represents the problem of poor join order. This pattern (**Pattern B**) is given by the following properties: (i) LOLEPOP of type JOIN (which means any type of JOIN method, e.g. NLJOIN, hash join (HSJOIN) and merge scan join (MSJOIN)); (ii) has a descendant (i.e., not necessarily immediately below) outer input stream of type JOIN; (ii) has a descendant inner input stream of type JOIN; (iii) the descendant outer input stream join is a Left Outer Join; (iv) descendant inner input stream join is a Left Outer Join. The recommendation for this pattern is to rewrite the query from the following structure (T1 LOJ T2) … JOIN … (T3 LOJ T4) to ((T1 LOJ T2).... JOIN ....T3) LOJ T4 as the rewritten query is more efficient. This optimization is now automatically done in DB2 but was found to be a limitation in early versions of DB2. This illustrates the usefulness of the tool in database optimizer development as well as supporting clients that use previous version of the DB2 system. We found QEPs matching this problem pattern in the real customer workload used in experiments, since the customer uses previous version of DB2. Figure 7 represents an example of the DBMS QEP that contains specified problem pattern. (Left outer join operators are prefixed in a QEP with ">" symbol, e.g, >HSJOIN and >NLJOIN.) This pattern is an example of the recursive problem pattern, since descendant outer and inner input stream of type LOJ do not have to be necessary immediate child of JOIN. (For instance, see LOLEPOP #5 and LOLEPOP #15) in Figure 7.

An alternate recommendation for this pattern, in case T1 = T3, is to materialize the column(s) from table T4 into table T1 and change the order of the operators from (T1 LOJ T2)… JOIN… (T1 LOJ T4) to ((T1 LOJ T2).... JOIN ....T1), eventually allowing to eliminate T4 as well as one instance of T1, because it had a unique key join to itself. This optimization is not automatically done in current version of DB2 optimizer.

The next pattern (**Pattern C**) represents the problem related to estimation of the execution cost by optimizer. This pattern is given by the following properties: (i) LOLEPOP of type index Scan

(IXSCAN) or table scan (TBSCAN) (ii) has cardinality smaller than 0.001; (iii) has a generic input stream of type Base Object (BASE OB); (iv) the generic input stream has cardinality bigger than 100000. The recommendation in this case is to create column group statistics (CGS) on equality local predicate columns and CGS on equality join predicate columns of the Base Object. Figure 8 represents an example of the DBMS query explain plan that contains specified problem pattern. With column group statistics, the optimizer can determine a better QEP and improve query performance. This is a common tuning recommendation when there is statistical correlation between column values associated with multiple equality local predicates or equality join predicates.

A user can also encounter a problem related to SORT spilling (**Pattern D**) given by the following properties: (i) LOLEPOP of type SORT; (ii) has an input stream immediately below with an I/O cost less than the I/O cost of the SORT. The recommendation may be to change the database memory configuration to increase sort memory if the number of QEPs containing this pattern is large enough to benefit the performance of many queries in the workload.

With expert or user provided patterns and recommendations in the knowledge base, OptImatch can iterate over all of the predetermined problem patterns in the knowledge base. Specifically, each problem pattern may, in such cases, be understood as being received from the knowledge base. If matched, OptImatch recalls and returns the recommendations corresponding to the particular problem pattern. Such a technique enables query plan checks to be routinized – a user can, with no particular knowledge or training, run a general test of all predetermined problem patterns against a given query workload.

## 3. EXPERIMENTAL STUDY

In this section, we present an experimental evaluation of our techniques. Our evaluation focuses on three objectives.

a) An evaluation of the effectiveness of our approach using real IBM customer query workload (1000 QEP files). (Section 3.2 and Section 3.3)

b) Scalability and performance over different problem characteristics: size of the query workload (Section 3.2.1), number of LOLEPOPs (Section 3.2.2) and number of recommendations in the knowledge base (Section 3.2.3).

c) A comparative study with manual search for patterns by experts, quantifying the benefits of our approach in terms of time and precision. (Section 3.3)

### 3.1 Setup

Our experiments were run on an Intel® Core™ i5-4330M machine with 2.80GHz processor and 8GB of memory.

For all the conducted experiments, we used three patterns created by IBM experts. Each pattern has associated recommendations for the user (stored in the knowledge base), providing a diagnosis of the artefact. The patterns used throughout the experimental study and their respective recommendations are as follows (detailed description of each pattern can be found in Section 2).

a) **Pattern #1** – **Pattern A** (Section 2.2) that represents a problem with recommendation related to indexing.

b) **Pattern #2** – **Pattern B** (Section 2.3) of a problem with a recommendation related to rewriting the query.

c) **Pattern #3** – **pattern C** (Section 2.3) that represents problem with a recommendation related to statistics for better cardinality (and consequently cost) estimation.

We measure the performance by computing system time for running OptImatch over the IBM customer query workload. We demonstrate the benefits of our framework by quantifying the running time against the size of the query workload (number of QEPs), number of LOLEPOPS (complexity of individual QEPs) and number of recommendations in the knowledge base.

### 3.2 Performance and Scalability

#### 3.2.1 Size of Query Workload

In the first experiment, we measure the performance of our tool by dividing the IBM customer query workload into ten buckets, each containing a different number of execution files. The first bucket contains 100 QEP files, and for each following bucket another 100 unique QEP files is added up to 1000. In other words, the distribution of the QEP files over buckets is [100, 200, ..., 1000].

The purpose of this experiment is to verify how efficient is our tool to search for portions of QEP files that match the prescribed pattern against different size of the query workload. The test was repeated six times (for each pattern), by dividing the QEP files into buckets randomly. (The average time is reported.)

Figure 9 reveals that time needed to compute the search increases *linearly* with the number of QEP files. The linear dependence allows our problem determination tool to scale well to large query workloads. Furthermore, the time to perform the search even for a large query workload with 1000 QEPs is less than 1min 10 seconds. Therefore, we can conclude that our tool allows for real time, on-the-fly search for patterns over diagnostic data.

The pattern #2 takes more time to be searched for than the others (around two times more). This is because the pattern #2 is more complex, as it contains descendant nodes. Therefore, recursion is used to analyze all LOLEPOPs inside the query explain plans which are fairly complex involving on average 100+ operators in the experimental workload.

#### 3.2.2 Number of LOLEPOPs

In the second experiment, we measured the performance of our system over QEPs with varying number of LOLEPOPs. We divided the IBM customer query workload into eleven buckets. The first bucket contains QEPs with the number of LOLEPOPs from 0 to 50, the second one from 50 to 100, and so on, until the last bucket that contains from 500 to 550 LOLEPOPs. (The maximum number of LOLEPOPs encountered in the workload was 550.) However, buckets 7-10 with the number of LOLEPOPS from 250 to 500 turned out to be empty, because the tested query workload contains only query explain plans with number of LOLEPOPS below 250 or above 500. Hence, as a consequence we report numbers for six buckets, 1-5 and 11. In other words, the distribution of the buckets is [0-50], [50-100], [100-150], [150-200], [200-250], and [500-550]. The number of pops is tied to the size of the explain file, the larger number of pops, the larger the size of the file.

The objective to run this experiment is to verify how efficient searching for patterns is as a function of number of pops. The experiment was repeated 6 times for each pattern, and the average time is reported. For each bucket, we report the average time in milliseconds, to analyze a single explain plan.

The results of this experiment are presented in Figure 10. As expected, the time spent to analyze QEPs increases as the number of LOLEPOPs increases. However, the time spent to analyze the QEPs increases in a linear fashion. Therefore, our system scales well for complex queries with a large number of LOLEPOPs. Moreover, even large and complex queries (with around 500

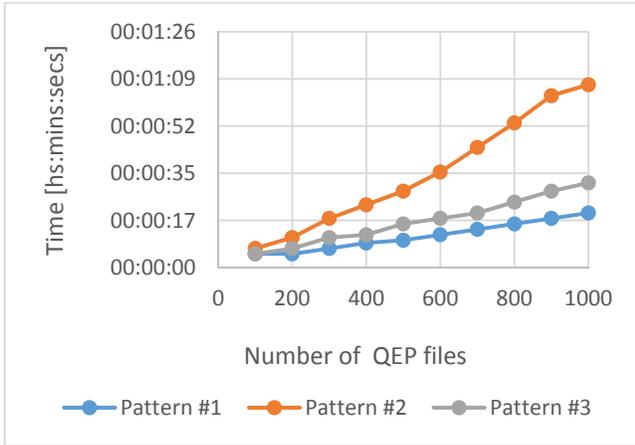

**Figure 9 Search time versus number of QEP files**

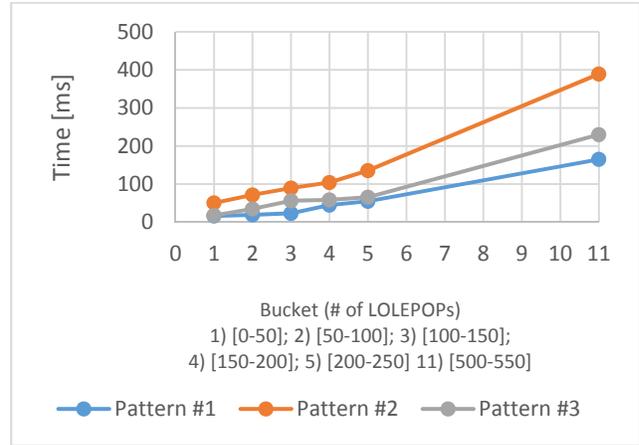

**Figure 10 Search time versus number of LOLEPOPs**

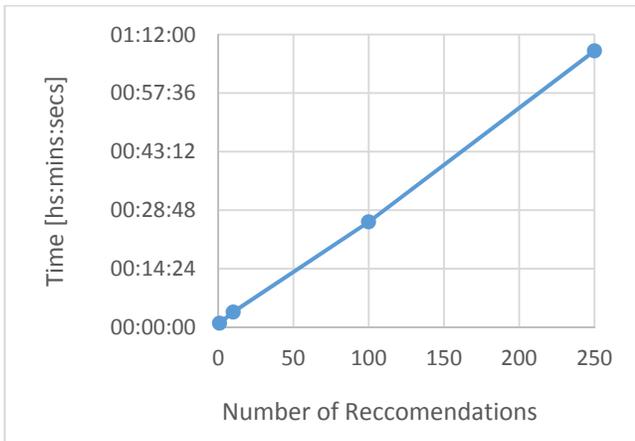

**Figure 11 Matching recommendations in knowledge base**

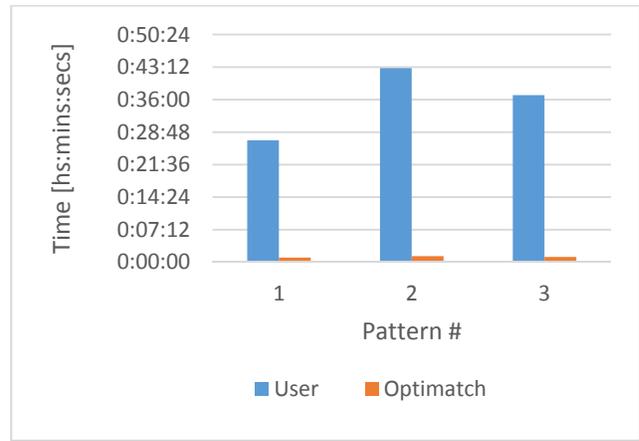

**Figure 12 Comparative user study**

LOLEPOPs) can be processed efficiently by our tool (less than 400 milliseconds).

*3.2.3 Number of Recommendations in Knowledge Base*
In the next experiment, we quantify the performance of our system against the number of recommendations in the knowledge base. We measure the running time to analyze 1000 QEP files against 1, 10, 100 and 250 recommendations in the knowledge base, respectively.

We perform this experiment to simulate the important use case for our system described in Section 2.3 to routinize query plan checks with expert provided predetermined problem patterns and corresponding recommendations against a given complex query workload. Instead of taking a specific problem pattern defined by the user, the system iterates over all of the predetermined problem patterns in the query plan knowledge base and provides matching solutions to known problems.

We report the results of this experiment in Figure 11. Our framework adapts well, with linear dependence over the number of recommendations in the knowledge base. The linear dependence allows our system to scale well to large knowledge databases. Our tool can process a 1000 query workload against 250 problem patterns and recommendations in around 70 minutes.

### 3.3 Comparative User Study
In the last experiment, we measure the time to perform pattern search both manually by experts and automatically with OptImatch. We also looked at the search quality. For each of the three patterns, we provide users the workload with 100 distinct QEP files. Out of 100 QEP files, 15, 12 and 18 QEP files match the three prescribed patterns #1, #2, and #3, respectively. Three experts participated in this experiment. We report the aggregated average statistics.

The purpose of this experiment is to quantify the benefits of our automatic approach against cumbersome manual search by experts that is prone to human error. The time comparison is shown in Figure 12. It can be observed that our tool drastically reduces the time to search for a pattern against even a sample of the query workload. (We perform this experiment over a sample of the query workload due to the limited time experts could spend to participate in the experiment.) Overall, our tool is around 40 times faster than the manual search by IBM experts. To simulate real world environment during the manual search for patterns experts were allowed to access the tools that they use in their daily problem determination tasks. An example of this includes the *grep* command-line utility for searching plain-text data sets for lines matching a regular expression.

**Table 1 Precision for manual search**

| Pattern # | #1 | #2 | #3 |
|---|---|---|---|
| Precision | 88% | 71% | 81% |

When we measure the running time for automatic search with our tool, we include the time both for specifying the pattern using graphical interface in our tool (on average around 60 seconds), as well as, performing the actual search by our system. Based, on this experiment, it can be inferred that manual search for a larger query workload (1000 queries) would take approximately 5 hours, whereas, with our tool this can be performed in around 2 minutes (around 150 times faster). Note that an automatic search pattern has to be specified only once by the user.

Last but not least, we report the quality of the search results in our comparative study. We measure the precision as the function of missed QEP files that contain the prescribed pattern. As predicted, manual search has been prone to human error. The precision for manual search by experts is on average 80%. Details are provided in Table 1. The common errors include misinterpreting information stored in the QEP file as well as formatting errors, e.g., using *grep* on operand value 0.001 while this information is represented in the QEP in either the decimal form or with an exponent as $10^{-3}$. Obviously, since our tool is fully automatic and immune to such differences, it provided 100% precision. Our system does not only perform significantly faster than a manual search but it also guarantees correctness. Often, high precision may be very important in problem determination analysis, as such, this experiment emphasizes another important benefit of OptImatch.

## 4. RELATED WORK

The SQL programming language is declarative in nature. Therefore, it is enough to specify what data we want to retrieve, without actually specifying how to get data. This is one of the main strengths of SQL, as it means that it should not make a difference to the query optimizer how a query is written as long as the different versions are semantically equivalent. However, in practice this is only partially true, as there is only a limited number of machine-generated query rewrites that a database optimizer can perform [7]. As the complexity of SQL grows, there is an increasing need to have tools help with performance problem determination.

Many different formalisms have been proposed in query optimization. We cite here only the most pertinent references. Join, sort and group by are at the heart of many database operations. The importance of these operators for query processing has been recognized very early on. Right from the beginning, the query optimizer of System R paid particular attention to interesting orders by keeping truck of indexes, ordered sets and pipelining operators throughout the process of query optimization, as described in Selinger et al. [12]. Within query plans, group-by, order-by and join operators can be accomplished either by a partition operation (such as by the use of a hash index), or by the use of an ordered tuple stream, as provided by a tree-index scan or by a sort operation (if appropriate indexes are not prescribed).

In Guravannavar et al. [9], authors explored the use of sorted sets for executing nested queries. The importance of sorted sets in query optimization has prompted the researchers to look beyond the sets that have been explicitly generated. In Szlichta et al. [15], authors show how to use relationship between sorted attributes discovered by reasoning over the physical schema via integrity constraints to avoid potentially expensive join operator. The inference system presented in follow-up work provides a formal way of reasoning about previously unknown or hidden sorted sets [16], [17]. Based on that work, many other optimization techniques from relational query processing can also be adapted to optimize group by, order by and case expressions [2], [18].

Optimization strategies described above hold the promise for good improvement. Their weakness, however, is that often the indexes, views and constraints that would be useful for optimization for a given database and workload is not explicitly available and there is only a limited number of types of query transformations that the optimizer can perform. Therefore, problem determination tools [21], [22] offer an alternative automated way to analyze QEPs and provide recommendations, such us re-write the query, create an index or materialized view or prescribe an integrity constraints. However, existing automatic tools for query performance problem determination do not provide the ability to perform workload analysis with flexible user defined patterns, as they lack the ability to impose proper structure on QEPs (as described in details in Section 1).

## 5. CONCLUSIONS

Query performance problem determination is a complex process. It is a tedious manual task that requires one to analyze a large number of QEPs that could span thousands of lines. It also necessitates a high level of skill and in-depth optimizer knowledge from users. Identification of even known issues is a very time and resource consuming and prone to human error.

To the best of our knowledge, we are the first to provide the system that performs interactive analysis in a structure manner of potentially a large number of QEPs in order to diagnosis and match optimizer problem patterns and retrieve corresponding recommendations that are provided by experts. Our semantic system combines and applies the benefits of RDF model and SPARQL query language in query performance problem determination and QEP analysis.

OptImatch is very well received and is proving to be very valuable in the IBM support of clients and database optimizer development organization by providing quick work around solutions through the use of a well-defined knowledge base.

Our methodology can certainly be applied to other general software determination problems (e.g., log data relating to network usage, security, or software compiling, as well as software debug data or sensor data relating to some physical system external to the system). Our framework can be applied to other general software problem determination, assuming that there exists automatically generated structured diagnostic information in the form of the graph than needs to be further analyzed by an expert or general user. This is the direction that we would like to explore in the future work.

## 6. ACKNOWLEDGMENTS

The authors would like to thank members of the DB2 optimizer development and support teams for their feedback and guidance through the development of the OptImatch system. Special thanks in particular to Shu Lin, Vincent Corvinelli and Manopalan Kandiah.

## 7. TRADEMARKS